# Portable QCD codes for Massively Parallel Processors

UKQCD collaboration, presented by Nick Stanford[a]

[a]Department of Physics, University of Edinburgh, Edinburgh EH9 3JZ, Scotland, UK

We present a new set of QCD codes in both message passing and data parallel versions. The message passing package used is PARMACS, although other packages may be used. Data parallel software is written in High Performance Fortran, an emerging standard based on Fortran 90. Software engineering methods have been applied to a physics application to create thoroughly tested and documented codes for the next generation of massively parallel supercomputers.

## 1. Introduction

UKQCD has for the last three years been reliant on a 64-node i860/transputer Meiko Surface "Maxwell", running with $\beta = 6.2$ and hypercubic lattice size of $24^3 \times 48$. A TMC 16K processor CM-200 Connection Machine "Euclid" has also been used (based at the Edinburgh Parallel Computing Centre) for development work and some production codes ($\beta = 6.0$ and lattice size of $16^3 \times 48$). The parallel codes have been written in C with CS-tools for message passing and i860 assembler (Meiko machine) and CM Slicewise Fortran (CM-200).

The types of code used at Edinburgh are

- **GAUGE**. Generates quenched gauge configurations using the Hybrid over-relaxed algorithm [1] with SU(2) subgroups. This consists of 1 Cabibbo-Marinari heatbath update, 5 over-relaxed updates and a reunitarisation, and uses the Wilson gauge action.

- **SOLVER**. Generates quark propagators from gauge configurations, using the SW action[2], and conjugate gradient or minimal residual solvers. Red-black and non-redblack decompositions have been used with unigrid methods.

- **CORRELATE**. Ties up quark propagators into particle propagators.

- **SMEAR**. Smears source and/or sink with Wuppertal/Jacobi smearing.

- **ANALYSE**. Turn particle propagators into physics results i.e. decay constants, masses, form factors etc.

## 2. The need for portable codes

The ABRC is purchasing an MPP system for physics research in the UK in 1993-1994 and UKQCD needs to be in a position to run codes on this system as soon as it is up and running. The existing codes have grown to the point where porting, expansion and extension are non-trivial.

We therefore need a new set of codes which are portable between many MPP platforms. This implies a need for both Message Passing and Data Parallel codes. With the high turn-over rate of students and Postdocs, and the often low experience of high performance computing and software engineering methods, these codes need to be fully and clearly documented for quick understanding, as do the methods used to create them and the rationale behind the design.

The nature of physics research is such that the functionality of the codes will change throughout their lifetime, with the implication that the codes must be easy to modify and extend.

## 3. System specification

### 3.1. Message Passing codes

These are written in standard Fortran 77 with PARMACS for message passing. This combination ensures a high success rate for compilation on new platforms. Fortran is used instead of C on the MPP machine because of the better vectorisation of compilers, comprehension by other scientists and ease of testing and debugging.

The PARMACS layer is isolated as far as possible, and has been designed to be easily changed to use different message passing packages. Schemes for the use of PVM, CHIMP and CS-tools have been designed, but not yet implemented.

### 3.2. Data parallel codes

These are written in subset High Performance Fortran (HPF) [3]. This is an emerging standard similar to Fortran-90 and CM Fortran, and although very few compilers are currently available it is likely that this subset (or something similar), which is designed to be quickly and easily implemented, will be available on new machines.

### 3.3. Workstation codes

Workstations are used to perform some operations in the system, such as analysis. These codes are written in C, exploiting its greater flexibility with text and file handling, memory access and data manipulation.

### 3.4. Required physics elements

In addition to the existing component types discussed previously we will need

- **HMC**. Generate dynamical gauge configurations with the SW action.
- **GAUGEFIX**. Gauge fix a given configuration (future option)

The codes are written using a library of common routines since so much code is shared between applications.

### 3.5. User interface

All parameters are input via plain text files with a free format (i.e. fields do not need to be in a fixed position) and labelling so that users do not need to have expert knowledge of the codes, and can modify run parameters as simply as possible without having to read a huge user manual. A common interface is to be used (i.e. one application writes out all batch scripts and runs the physics codes) again so that new and/or occasional users do not have to waste time relearning the system for a different application.

## 4. Design methodology

As there is more than one person working on the project care needs to be taken when defining interfaces between modules. Documentation must be written in parallel with design work so that information and understanding can be shared between project members in a readable format. This is the area where the creation of most physics codes suffers, physicists are not typically taught any engineering techniques in the early years of their careers, and attack such a large problem as this with a piecewise rather than global view, not giving sufficient consideration to interfacing of components and division of labour.

We have adopted a modified version of the 'Waterfall' method [4] of software design as shown in figure 1 in conjunction with Yourdon methodology [5] for structured analysis.

## 5. Current position

- **GAUGE**. Working fully in MP/HPF versions. Runs on Sun4, DEC-ALPHA, CM-200, Cray YMP-8, INTEL iPSC-860 and MEIKO CS-1-860. Vectorisation is being improved at the moment.

- **SOLVER**. Working fully in HPF, and without Clover in MP, runs on the same machines as above. This code vectorises extremely well.

  This solver is due to appear in the GENESIS and PARKBENCH benchmark suites.

- **HMC**. Written by Mike Peardon, using the existing library of routines. Working in HPF with SW action. Only running on CM-200 at present. MP version is planned for the next few months.

- **GAUGEFIX**. An empty slot has been defined for this application, although no design work has yet been planned.

- **SMEAR, CORRELATE, ANALYSE**. Not yet designed, work is planned to start after Lattice '93.

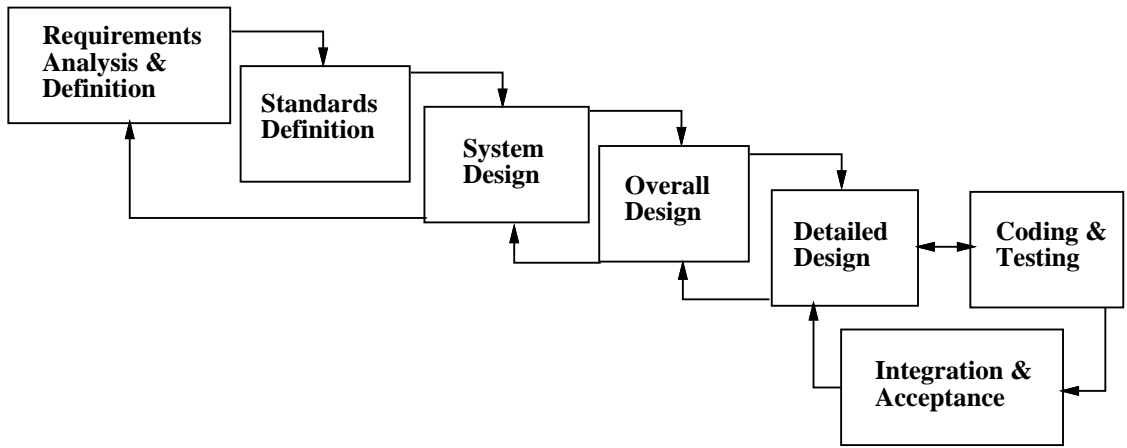

Figure 1. The Waterfall method of Software Engineering.

## 6. Conclusions

In the spring of 1994, UKQCD will have a portable set of QCD codes with a fully documented and tested library of routines for further expansion and extension. The rigid testing of the codes adds greater confidence in the accuracy of physics results extracted, a common problem with massively parallel codes.

Applying software engineering methods to a physics research project of this magnitude (a new approach for most research groups) need not be difficult, and has many benefits.

## 7. Acknowledgements

I wish to acknowledge the financial support of the SERC and the University of Edinburgh and the guidance of Dr. Richard Kenway and Dr. Stephen Booth. The Meiko i860 Computing Surface used is supported by SERC grant $GR/G$32779, Meiko Ltd., and the University of Edinburgh. The Thinking Machines CM-200 is supported by SERC, Scottish Enterprise and the Information Systems Committee of the UFC. Other support has been provided by SERC grant $GR/H$01069.

## REFERENCES

1. A. D. Kennedy, Proceedings of the 1992 Symposium on Lattice Field Theory, Nucl. Phys. B Proc. Suppl. 30 (1993).
2. B. Sheikholeslami and R. Wohlert, Nucl. Phys. B259 (1985) 572–596.
3. High Performance Fortran Language Specification Version 1.0, High Performance Fortran Forum, Rice University, Houston Texas (1993).
4. Ian Sommerville, Software Engineering, Addison-Wesley (1992).
5. Edward Yourdon, Modern Structured Analysis, Prentice Hall international (1989).